\documentclass[twocolumn,pre,superscriptaddress]{revtex4}

\usepackage{dcolumn}
\usepackage{amsmath}
\usepackage{url}
\usepackage{graphicx}

\begin{document}

% Macros
\renewcommand{\d}{\mathrm{d}}
\newcommand{\ii}{\mathrm{i}}
\newcommand{\Ord}{\mathrm{O}}
\newcommand{\e}{\mathrm{e}}
\newcommand{\half}{\mbox{$\frac12$}}
\newcommand{\set}[1]{\lbrace#1\rbrace}
\newcommand{\av}[1]{\langle#1\rangle}
\newcommand{\etal}{{\it{}et~al.}}
\newcommand{\defn}{\textit}
\newcommand{\stirling}[2]{\biggl\lbrace{\!{#1\atop#2}\!}\biggr\rbrace}
\newcommand{\bsy}{\boldsymbol}
\newcommand{\beq}{\begin{equation}}
\newcommand{\eeq}{\end{equation}}

% Style parameters
\newlength{\figurewidth}
\setlength{\figurewidth}{0.95\columnwidth}
\setlength{\parskip}{0pt}
\setlength{\tabcolsep}{6pt}
\setlength{\arraycolsep}{2pt}

\title{Hypergraph topological quantities for tagged social networks}
\author{Vinko Zlati\'{c}\thanks{vzlatic@irb.hr}}
\affiliation{CNR-INFM Centro SMC Dipartimento di Fisica, Universit\`a di Roma ``Sapienza'' P.le Moro 5, 00185 Roma, Italy }
\affiliation{Theoretical Physics Division, Rudjer Bo\v{s}kovi\'{c} Institute, P.O.Box 180, HR-10002 Zagreb, Croatia}
\author{Gourab Ghoshal}
\affiliation{Department of Physics and Michigan Center for Theoretical Physics, University of Michigan, Ann Arbor, Michigan 48109, USA}
\author{Guido Caldarelli\thanks{Guido.Caldarelli@roma1.infn.it}}
\affiliation{CNR-INFM Centro SMC Dipartimento di Fisica, Universit\`a di Roma ``Sapienza'' P.le Moro 5, 00185 Roma, Italy }
\affiliation{Linkalab, Complex Systems Computational Lab. 09100 Cagliari Italy}

\begin{abstract}

Recent years have witnessed the emergence of a new class of social networks, that require us to move beyond previously employed representations of complex graph structures. A notable example is that of the folksonomy,  an online process where users collaboratively employ tags to resources to impart structure to an otherwise undifferentiated database.  In a recent paper~\cite{Gourab09} we proposed a mathematical model that represents these structures as tripartite hypergraphs and defined basic  topological quantities of interest. In this paper we extend our model by defining additional quantities such as edge distributions, vertex similarity and correlations as well as clustering.  We then empirically measure these quantities on two real life folksonomies, the popular online photo sharing site Flickr and the bookmarking site CiteULike. We find that these systems share similar qualitative features with the majority of complex networks that have been previously studied. We prop!
 ose that the quantities and methodology described here can be used as a standard tool in measuring the structure of tagged networks. 
\end{abstract}
\maketitle
\section{Introduction}

Over the past decade, the recognition of complex networks as a useful and versatile mathematical representation of various real world systems, has led to a huge volume of work, studying its topological and dynamical properties~\cite{JMBO01,FFF99, BCLM03, WF94,Newman00,Albert01}.  A variety of models have been proposed, ranging from those describing simple undirected graphs, a basic representation of a communication network for example, to more complicated bipartite networks representing collaboration networks such as board of directors in a company, or movie actors---see~\cite{NewmanRev,DM04,Caldarelli07}.

However, the advent of Web 2.0 and its associated new forms of user-driven content have led to new social systems that cannot be adequately described by existing models. One such example is related to a phenomenon known as \emph{folksonomy}~\cite{ciro,PINTS}.  In this process, users collaboratively create and manage \emph{tags}  to categorize and annotate data.  Unlike traditional forms of data indexing, where administrators of a particular web-page maintain and categorize the content, in a folksonomy, both creators and consumers are free to participate in the process. Instead of a controlled set of keywords, tagging networks consist of a user generated taxonomy.  

Consider the example of the popular file-sharing database known as Flickr.  In this website, users can create an account and upload their personal photos. In addition to uploading photos, they are free to give them a short text description using tags.  These photos (in most cases) can then be viewed by other users, who in turn can assign additional tags to the photo depending on their preferences, and so the process continues.  There are also a number of other websites of a similar nature, but dealing with different resources.  In the website CiteUlike, for example, users upload and assign tags to academic papers as opposed to photographs.

Roughly speaking, tagged networks can be divided into two categories. In the first case, users are presented with a variety of available key words, which they can then freely employ to resources of their choice. Although this represents a degree of control in the set of tags that are available to users, the mechanism by which this control arises is still decentralized. In Flickr for example,  when a user uploads a photograph and gives it a short text description, that description or tag is always public, which is to say that anyone visiting the site can see the full set of tags describing the photograph. 
This serves a number of functions. On the one hand it prevents the practice of redundant tagging, since once a particular tag has been applied to a resource, one is not allowed to retag the item with the same description; on the other hand it also provides new users with a previously employed set of popular tags which they can then use on their own photographs. Finally, if none of the previously employed tags are appropriate to newly uploaded resources, then users are forced to supply sufficiently different descriptions. In this way the set of keywords present in the network represent a reasonably well organized  and diverse set.  In other websites such as Citeulike, tags are not always public, and this process of decentralized control is not present. Consequently, this may give rise to vastly different statistical properties.

Some attempts have been made recently, to try and describe these tagging systems. Among them, people have tried to model them as simple unipartite and bipartite graphs, as well as simplified forms of tripartite graphs~\cite{Sapienza_folkpaper, Ausloos_Lambiotte_2006, PFPDV_2008}.  In addition there have been a number of studies focussing particularly on the tags such as the definition of communities~\cite{Specia07}, clustering~\cite{Capocci08} and global measures such as PageRank~\cite{Hotho06}.

The key thing to note however, is that unlike in a simple network merely consisting of vertices, and edges describing the association between them, in a tagging network the fundamental building block is a triple consisting of a \emph{user}, a \emph{resource} that the user uploads and finally a \emph{tag} that the user employs to describe the resource.  A complete representation of such folksonomy data must capture this three-way relationship, and this leads us to consider hypergraphs.

A hypergraph is a generalization of a regular graph in the sense that an edge can connect multiple vertices.  So unlike in a regular graph, where an edge connects two vertices, in a hypergraph a hyperedge is a collection of arbitrary number of vertices.  These vertices can be of the same or different types, and hyperedges can vary in the number of vertices they connect.  This fits in quite nicely with how tagging networks are organized. By representing a triple, as a hyperedge, one can conveniently preserve the structure of the network and examine its properties in its entirety.

In a previous paper~\cite{Gourab09} we defined a mathematical null model that represents these folksonomies as random tripartite hypergraphs and defined some basic topological quantities of interest such as the degree distribution, and component structure. In addition we calculated a number of properties of the model in the limit of large system size. In this paper, based on the hypergraph representation, we define a number of other useful topological features, such as the edge distribution, hyperedge distribution, vertex similarity, distance measures and the clustering coefficient as well as a simple definition of community structure based on the similarity between vertices. We then measure these quantities on datasets gathered from two real folksonomies, Flickr~\cite{dataF} and CiteULike~\cite{dataC}. We find that these networks share a number of qualitative features with previously studied social networks.

\begin{figure}[t]
\centerline{%\rotatebox{-90}
{\resizebox{0.25\textwidth}{!}
{\includegraphics{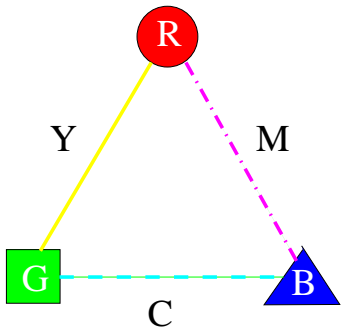}}}}
\caption{\label{Fig:HyperEdge} A hyperedge representing the fundamental building block in our network. Each hyperedge consists of three types of vertices, red (circles), green (squares) and blue(triangles). In addition the regular edges are also colored according to the types of vertices they connect.  In relation to folksonomies, the circles represent the users, the squares the resources and the triangles the tags.}\end{figure}

\section{Tripartite hypergraphs}

We begin our analysis of the folksonomies, by first defining the representation that we will be using.  We represent the network as tripartite graphs consisting of three different types of vertices, which we will refer to as red, green and blue.  For the purposes of our study, red will represent resources, blue tags and green users, however the colors themselves are secular as to what they represent. The edges represent three-way hyperedges that each connect exactly one red, one green and one blue vertex. 
In addition we can also color the regular edges, depending on the types of vertices they connect. For example the edge connecting a blue and green vertex is cyan (since blue and green combine to form cyan in the visual spectrum). Similarly the other edges are colored yellow (red-green) and magenta (red-blue). This classification of different regular edges allows us to measure quantities such as, the number of hyperedges a given regular edge participates in and so on. A visual illustration of this is shown in Fig.~\ref{Fig:HyperEdge}.

To couch this in the language of graph theory, our representation corresponds to the case of a tripartite hypergraph $\bsy{G}=(\bsy{V},\bsy{H})$ which can be defined as a 
pair of sets $\bsy{V}$ and $\bsy{H}$, that satisfy 
the following conditions:{\em (i)} the set $\bsy{V}=\lbrace \bsy{V}_r,\bsy{V}_g,\bsy{V}_b 
| \bsy{V}_i\cap\bsy{V}_j=\oslash\rbrace$ 
is formed by the union of three disjoint sets, and{\em (ii)} the set 
$\bsy{H}\subset\lbrace (v_r\in\bsy{V}_r,v_g\in\bsy{V}_g,v_b\in\bsy{V}_b)\rbrace$ 
of hyperedges is a triangle connecting elements of these three sets.  

In~\cite{Gourab09} we investigated number of basic properties of such hypergraphs, such as the tripartite analog of the vertex degree, component structures  and projections of the network into the space of bipartite and unipartite graphs. In addition we defined a random graph model, related to a version of the configuration model for regular graphs~\cite{Molloy,Newman2001} and calculated these properties exactly in the limit of large graph size. One of the assumptions made in the model was that the hypergraphs were locally treelike, in the sense that there were a trivial number of short-range loops connecting vertices. In real folksonomies however, this assumption is not strictly true, and to reflect this we extend our model by defining a number of other properties of interest to examine the loop structure as well as correlations in the network.  

In particular we measure the following quantities:

\begin{itemize}
\item  {\em edge degrees:} defined as the number of hyperedges that a regular edge participates in. For example, a magenta edge connecting red and blue vertices might participate in a triple with a number of other green vertices. In the language of folksonomies, this represents the number of resources that a user has described with the same tag.	
\item {\em clustering:} defined as the degree of overlap between the different hyperedges that a vertex participates in.
\item {\em vertex-vertex distance:} defined as the shortest paths between two nodes that are reachable along hyperedges, as well as via colored regular edges.
\item {\em community structure:} defined on the basis of vertex similarity between nodes of the same type. 
\end{itemize}

\subsection{Degrees}

\begin{figure}[t]
\centerline{%\rotatebox{-90}
{\resizebox{0.5\textwidth}{!}
{\includegraphics{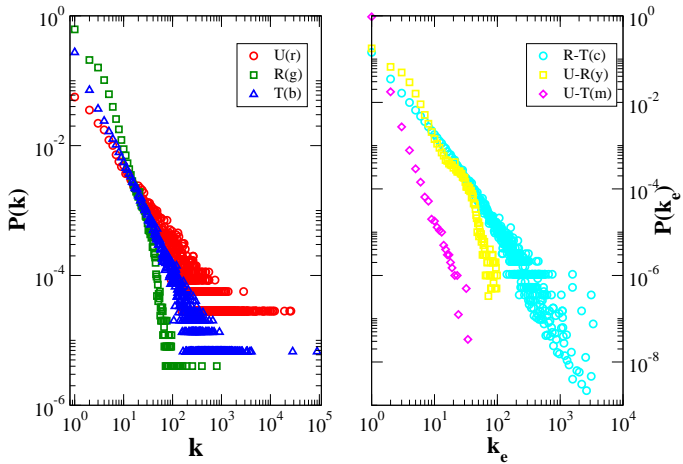}}
}}
\caption{(color online)
The two types of hyperdegree distributions found in a subset of the website CiteULike. Left panel:  The vertex degree distributions for users, tags and resources. Right panel: The degree distribution for the various edge types.}
\label{Fig:HyperDistroCite}
\end{figure}

There are a number of options available to us when defining the degree of a vertex or edge in a tripartite graph. The simplest and most reasonable choice for a vertex is to count its degree as the number of hyperedges it participates in. Thus a red vertex that connects to four hyperedges has degree four. The same applies to vertices of different colors. If there are $H$ hyperedges in the network, and $N_r$ red, $N_b$ blue and $N_g$ green vertices, then the mean degree of each vertex is fixed by the condition,
\begin{equation}
N_r c_r = N_b c_b = N_g c_g,
\label{eq:cond1}
\end{equation}
where $c_r$ represents the mean degree of red vertices, with $c_b$ and $c_g$ the corresponding quantities for blue and green. (This follows from the fact that each hyperedge consists of a single red, green and blue vertex.)

Just as in the case of regular graphs, we can define a degree distribution for each of the colors. We define $P(k_r)$ to be the fraction of red nodes in the network with hyperdegree $k_r$, as well as  $P(k_b)$ and $P(k_g)$ corresponding to blue and green respectively. These probability distributions satisfy the usual sum rules,
\begin{equation}
\sum_{k_r=0}^\infty P(k_r) = \sum_{k_g=0}^\infty P(k_g)
  = \sum_{k_b=0}^\infty P(k_b) = 1,
\label{eq:sumrule1}
\end{equation}
with $c_r = \sum_{k_r} k_r P(k_r)$ and similarly for the other two colors.

\begin{figure}[t]
\centerline{%\rotatebox{-90}
{\resizebox{0.5\textwidth}{!}
{\includegraphics{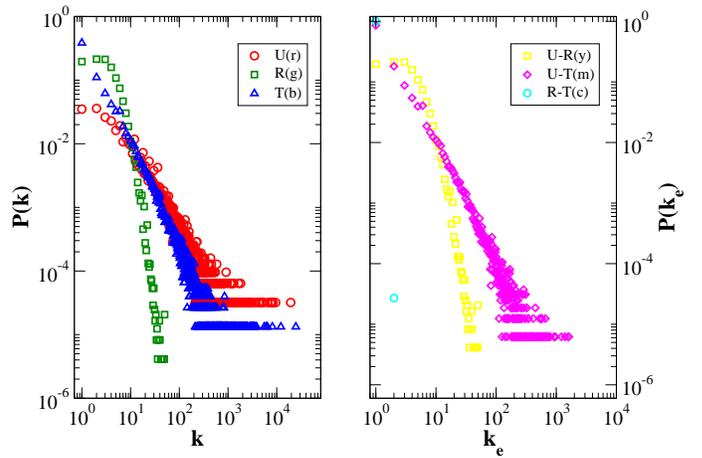}}}}
\caption{(color online)
The two types of hyperdegree distributions found in a subset of the website Flickr. Left panel:  The vertex degree distributions for users, tags and resources. Right panel: The degree distribution for the various edge types. The near absence of data points for the cyan edge (resource-tag) is related to the fact that tags in Flickr are public and this prevents redundant tagging---that is the application of the same tag to the same resource by multiple users.}
\label{Fig:HyperDistroFlickr}
\end{figure}

In addition to the degrees of vertices, we can also define corresponding quantities for regular edges. Say there are $H_y$ number of  yellow edges, then we define the degree of one of these edges as $k_y$---the number of different hyperedges it contributes to. We can think of these edges as representing pairs of vertices, such that in the context of folksonomies, the degree $k_y$ corresponds to the number of different tags that a given user applies to a particular resource. The quantities $k_m$ and $k_c$ represent the same for the other two types of edges. In exactly the same way as for the vertices, we can define edge degree distributions thus,
\begin{equation}
\sum_{k_c=0}^\infty P(k_c) = \sum_{k_y=0}^\infty P(k_y),
  = \sum_{k=0}^\infty P(k_m) = 1.
\label{eq:sumrule2}
\end{equation}
 with the mean degree $c$ of each edge fixed by the condition,
 \begin{equation}
 H_c c_c = H_y c_y = H_m c_m,
 \label{eq:cond2}
 \end{equation}
where $H = H_c + H_y + H_m$.

We measure these two different quantities on datasets on our two example folksnomies, CiteUlike and Flickr. On the left side of Figs.~\ref{Fig:HyperDistroCite} and~\ref{Fig:HyperDistroFlickr} 
we show the vertex degree distribution for both websites. As is fairly common for most social networks both of these show a fat-tailed distribution. On the right hand side of each figure we show the edge degree distribution.  Once again these distributions are typically right skewed. A notable difference is the distribution of common users for a given resource-tag pair in Flickr (the cyan edge).  As discussed in the introduction, this is related to the different tagging schemes in the two networks. Note that the phenomena of multiple users applying the same tag to the same resource is representative of redundant tagging or some sort of spam. Since in CiteUlike, the tags applied by a user to a resource is not always visible to other users of the website, the incidence of users applying the same description to a paper is much higher. In Flickr, however the tags are public, and once a tag is applied to a photograph no one else is allowed to employ the same tag to that photo---t!
 hus the near absence of any data points in the distribution of cyan edges.

\begin{figure}[t]
\centerline{%\rotatebox{-90}
{\resizebox{0.5\textwidth}{!}
{\includegraphics{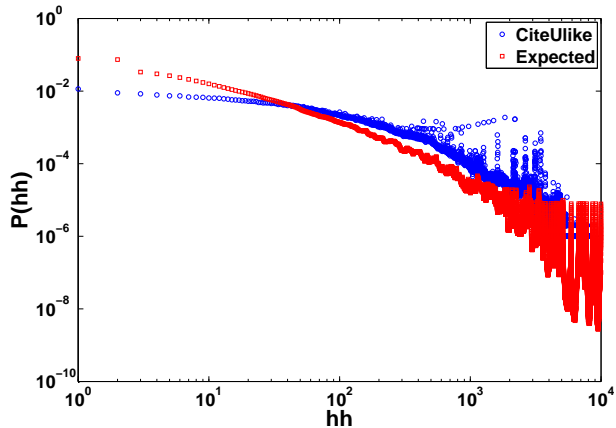}}}}
\caption{ 
The distribution of hyperedge neighbors $P(hh)$ as measured from the dataset in CiteULike (the spikes in the data are related to spamming), as well as the theoretical values computed from equation (\protect\ref{Hedgeexp}). }
\label{Fig:HEN} 
\end{figure}

Apart from the individual vertices and edges, we can also consider a hyperedge or triple as a basic unit and measure the number of other hyperedges, say $hh$, that are connected to it  via any of its individual constituents. In the spirit of regular unipartite graphs, one can loosely think of this as a measure of degree-degree correlations.

For a given hyperedge this quantity can be easily computed from the individual degree of each of its constituent vertices and edges in the following manner,
\beq
\label{HedgeCorr}
hh=k_r+k_g+k_b-k_c-k_m-k_y,
\eeq
where the indices represent the color of the different vertices and edges. So say for example, our network just consists of two hyperedges that share a common red vertex and we look at one of the triangles and examine the number of other hyperedges it connects to. Each of the blue and green vertices have degree one, all the edges have degree one, whereas the red vertex has degree two (since it is part of two hyperedges), then Eqn.~(\protect\ref{HedgeCorr}), correctly tells us that our hyperedge has degree one.
 
\begin{figure}[t]
\centerline{%\rotatebox{-90}
{\resizebox{0.5\textwidth}{!}
{\includegraphics{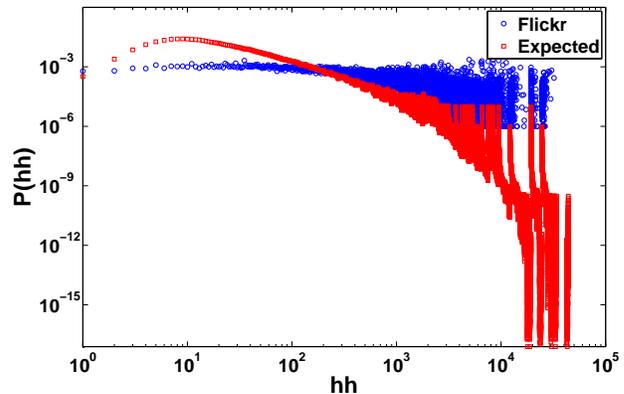}}}}
\caption{ 
The distribution of hyperedge neighbors $P(hh)$ as measured from the dataset in Flickr as well as the
theoretical values computed from equation (\protect\ref{Hedgeexp}). It is evident that the assumption of statistical independence between vertex and edge degrees is  not appropriate in this case. }
\label{Fig:HEN2} 
\end{figure}

In the same way as the vertex and edge degrees we can also define and calculate the distribution for the number of hyperedge neighbors of a given hyperedge. Let $P(hh)$ represent the fraction of hyperedges in the network that are connected to exactly $hh$ other hyperedges in the sense described above. Assuming that there are no correlations between the degrees of the vertices and the edges,
\begin{eqnarray}\label{Hedgeexp}
P(hh)&=&\sum_{\substack{k_{r},k_{g},k_{b}, \\ 
k_{c},k_{m},k_{y}}}P(k_{r})P(k_{g})P(k_{b})P(k_{c})P(k_{m})P(k_{y})\nonumber\\
&&\cdot\Theta(k_r-k_m-k_y)\Theta(k_g-k_c-k_y)\nonumber\\
&&\cdot\Theta(k_b-k_m-k_c)\delta_{hh,k_r+k_g+k_b-k_c-k_m-k_y},
\end{eqnarray}
where $\Theta(x)$ represents the Heaviside step function  and $\delta_{x,y}$ is the Kronecker delta.
In Figures~\ref{Fig:HEN} and~\ref{Fig:HEN2} we show the distribution of measured hyperedge degrees from the data sets  (blue circles) as well as the values predicted by Eq.~(\protect\ref{Hedgeexp}) (red circles). As is clearly visible the agreement between the two curves for both Flickr and CiteUlike is at great variance, thus suggesting that the degree of the vertices as well as the edges are correlated in some fashion and cannot be treated independently.

\subsection{Clustering}

As is well known, many networks show a high degree of clustering or transitivity---the tendency of two neighbors of a given vertex to also be neighbors of each other---thus forming triangles of connections. The average of the probability of such types of connections is called the clustering coefficient. It is instructive to determine if this effect is predominant in folksonomies. 

Once again there are a number of ways to define clustering in tripartite graphs. For example, one can project the graph onto the space of a vertex of a particular color (say that of users~\cite{Gourab09}), and use the standard measure of clustering. However, as discussed before, our aim is to keep the tripartite structure of our network intact, and therefore we will define an analog of clustering that takes into account the full three-way relationship between the vertices. 

As motivation for our definition of clustering, consider a red vertex (a user), that is connected to three 
hyperedges.  If the graph was locally treelike, then this would imply that the red vertex connects to three blue and three green neighbors. However it is possible  that some of its neighbors might be common to more than one hyperedge, thus the the number of blue and green neighbors could be less than three. One example is if a user assigns three tags to the same resource, then it has three tags as neighbors and only one resource.  Thus one can think of this measure of overlap between different hyperedges as a close analog to clustering for regular graphs, in the sense that it is a metric for the deviation of the network from being treelike. 

In order to quantify this measure, we first define the coordination number $z$ for a given vertex, as the number of immediate neighbors of any color that are connected to it via regular edges (this is just the standard definition of degree for regular graphs).  For a vertex with $k$ hyperedges  one can define upper and lower bounds for the coordination number. If there were no overlap, that is to to say, the vertex shares no common neighbors between its $k$ hyperedges, then the maximal coordination number $z_{max}$ is equal to $2k$, since it is connected to two other vertices via each hyperedge. One can show that in the case of maximum overlap, the corresponding expression for $z_{min}$ is  
$z_{min}(k)=2n$ for $n(n-1)  < k \leq n^2$ and $z_{min}=2n+1$ for $n^2 < k \leq n(n+1)$, with $n$ some integer. 

\begin{figure}[t]
\centerline{%\rotatebox{-90}
{\resizebox{0.45\textwidth}{!}
{\includegraphics{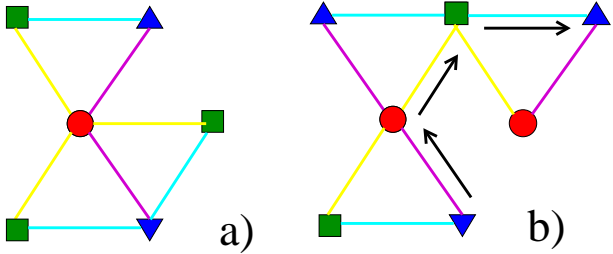}}}}
\caption{ 
In a) the central red node has $D_h=0.5$ as per Eq.~(\protect\ref{Hdensity}), because $z_{max} = 6$ and $z_{min}=4$. In b) a path of distance $3$ (marked by arrows) between two tags via a tag-user, user-resource and resource-tag path.}
\label{Fig:Explanations} 
\end{figure}
 
Based on the coordination number defined above for a vertex of degree $k$, we define a local measure of overlap or clustering, the \emph{hyperedge density} $D_h(k)$  thus:
\begin{equation}\label{Hdensity}
D_h(k)=\frac{z_{max}-z}{z_{max}-z_{min}}.
\end{equation}
It can be immediately seen that if a vertex does not share any common neighbors between its hyperedges, then $z =  z_{max}$, and the hyperedge density vanishes, which the means the neighborhood of the vertex is locally treelike. In the case of maximum overlap,  $z = z_{min}$, and the ratio is then $D_h = 1$. In Fig.~\ref{Fig:Explanations}a. we show an example analysis of the hyperedge density. 

\begin{figure}[t]
\centerline{%\rotatebox{-90}
{\resizebox{0.5\textwidth}{!}
{\includegraphics{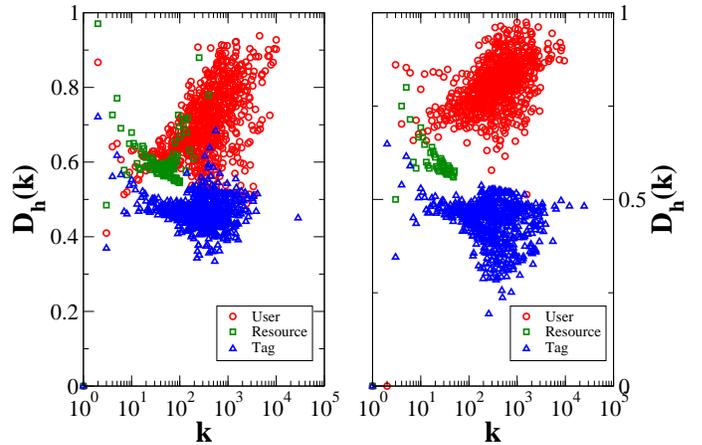}}}}
\caption{The average hyperedge density $D_h(k)$ as a function of the degree $k$ of a vertex. 
On the left panel, the hyperedge density for different vertex categories in the CiteULike dataset. On 
the right panel, the same quantities for Flickr.  For both datasets the users tend to form increasingly dense connections with increasing $k$.  The density for resources seem to fall off with increasing $k$, except in CiteULike, where there is an interesting turning point around $k = 100$.}
\label{Fig:DensityHyperEdge} 
\end{figure}

In Fig.~\ref{Fig:DensityHyperEdge} we show the measurement of the average hyperedge density as a function of a nodes degree for both our example websites.  The plot shows that both CiteULike and Flickr share a high incidence of overlap between its hyperedges. In fact the value of $D_h$ is generally always larger than $0.5$, which is to suggest connections of the type shown in Fig.~\ref{Fig:Explanations}a. are fairly common. For both types of networks, the hyperedge density of users is significantly larger than that of resources or tags.  There is possibly a fairly simple explanation for this. In Flickr for example, users typically upload photographs in sets and then apply descriptive tags to the same set. Thus many different resources share similar tags associated with the same user. Therefore although a user might participate in a number of hyperedges a majority of them are associated with either a common resource that has been assigned multiple tags, or a common tag that h!
 as been used to describe a number of resources by the same user. The lower hyperedge density values for individual tags imply that they are employed by a large variety of different user-resource pairs. In the case of Flickr, this might be reflective of the more diverse and ordered set of tags that arise due to the decentralized control described in the introduction.

\subsection{Vertex-Vertex Distance}

Another important quantity of interest is the average distance between a pair of vertices in a graph. This is important for a number of reasons. One application is related to searching for resources in a network. In CiteUlike for example, a user might be interested in looking for a particular paper. In order to do so one would have to \emph{surf}  the hypergraph, through the various hyperedges. In some sense the efficacy of the search is related to how far apart vertices of different types are in the network. For example, one might find that surfing on the network of tags would lead one to a resource in much faster time, than if one were to look through the list of users. The same considerations apply to automated \emph{web crawlers} that crawl through websites to perform directed searches, or to create indexes for later search. The knowledge of the distance between various types of vertices in the network can lead to more efficient paths being chosen and thus more effective!
  search schemes.

In the case of tripartite-hypergraphs the shortest paths between different 
vertices can be defined as the minimal number of hyperedges which connect 
those vertices. This definition follows from the definition of shortest paths 
in ordinary graphs. The flow of information through the hyperedges can be simply 
described as a hopping process along vertices sharing a common hyperedge. 
In addition to this it is also interesting to measure the paths through the differently colored regular edges. 
As mentioned before this might help in defining an efficient hopping scheme for an automated crawler which could try differently colored paths depending on which one is closest to a desired target at each step of the crawl. An example of the distance between two vertices (tags in this case) based on the hyperedges and the regular colored edges is shown in Fig.~\ref{Fig:Explanations}b.

%@Guido and Vinko. 
%Do we need this? It also is a bit unclear what you are trying to say here.

%We can also claim that in a random tripartite-hypergraph an average shortest 
%path should behave as:
%\beq\label{ShortPath}
%\langle l \rangle \sim \frac{\log{N}}{\log{\langle z \rangle}},
%\eeq
%where $\langle z\rangle$ represents the average coordination number and $N$ the 
%size of the network. This can be shown as follows: starting from a vertex $i$ 
%the shortest paths to other vertices passes through all the adjacent edges while 
%the edges opposite to the vertex will not contribute in any way to the shortest 
%paths, and therefore the result for the regular random graph of size $N$ and 
%average degree $\langle z\rangle$ is also valid in this case.

\begin{figure}[t]
\centerline{%\rotatebox{-90}
{\resizebox{0.5\textwidth}{!}
{\includegraphics{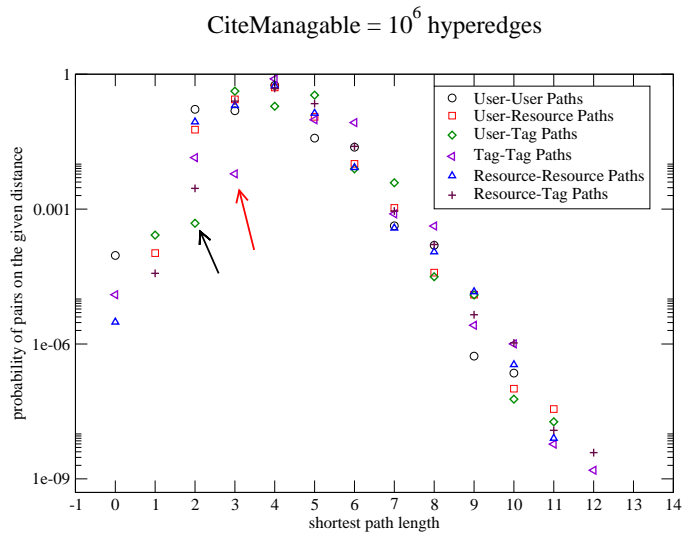}}}}
\caption{The distribution of paths between different types of vertices, implying that the network as a whole exhibits the so called  small-world effect.
The black arrow shows that the number of user-tag paths of 
distance $2$ are comparatively few. This means that there are very few resources that function as direct bridges between a user and a tag. The red arrow shows that are fewer paths between tags of length 3, as compared to those of length $2$ and $4$. This implies that the number of bridging hyperedges where   two tags are connected via a user and then a resource (as shown in Fig.~\ref{Fig:Explanations}b.) are rare in this network.}
\label{Fig:Paths} 
\end{figure}

We took a subset of the data from the website CiteUlike (denoted CiteManageable), and measured the average distance between vertices of the same type as well those of different types. The results are presented in Fig.~\ref{Fig:Paths}. In all cases, it seems that the average distance peaks around paths of length four, which suggests that the network as a whole exhibits the small-world effect~\cite{Strogatz_Watts98} also present in a variety of other networks. 

% No need for this I think. Seems out of place.

%From the distribution of paths it is possible to verify that in the  
%CiteULike dataset is present a giant connected component. 
%In~\cite{Gourab09} we have shown that general criteria for the formation 
%of giant connected component is
%\beq\label{Formation}
%\frac{\langle k_r\rangle}{\langle k_r^2\rangle}+\frac{\langle k_g\rangle}{\langle k_g^2\rangle}+\frac{\langle k_b\rangle}{\langle k_b^2\rangle}<2.
%\eeq
%This criterion reminds on the general criterion of 
%Molloy and Reed~\cite{Molloy} for the unipartite random graphs, 
%and is satisfied in the case of CiteULike data.

\subsection{Community Structure}

A question of particular importance, is to examine whether our example folksonomies exhibit
community structure---the tendency of the network to divide naturally into groups of nodes with dense connections within groups and sparser connections between groups.  For example there might be different groups of users in the network that share commonality with themselves owing to similar tastes in content---in Flickr for example users who share an interest in pictures of art.  Or one might find groups of tags that occur together many number of times---say \emph{differential}, \emph{equation}, \emph{series expansion}, when describing physics papers in the website CiteULike. 

\begin{figure*}[t]
\centerline{%\rotatebox{-90}
{\resizebox{1\textwidth}{!}
{\includegraphics{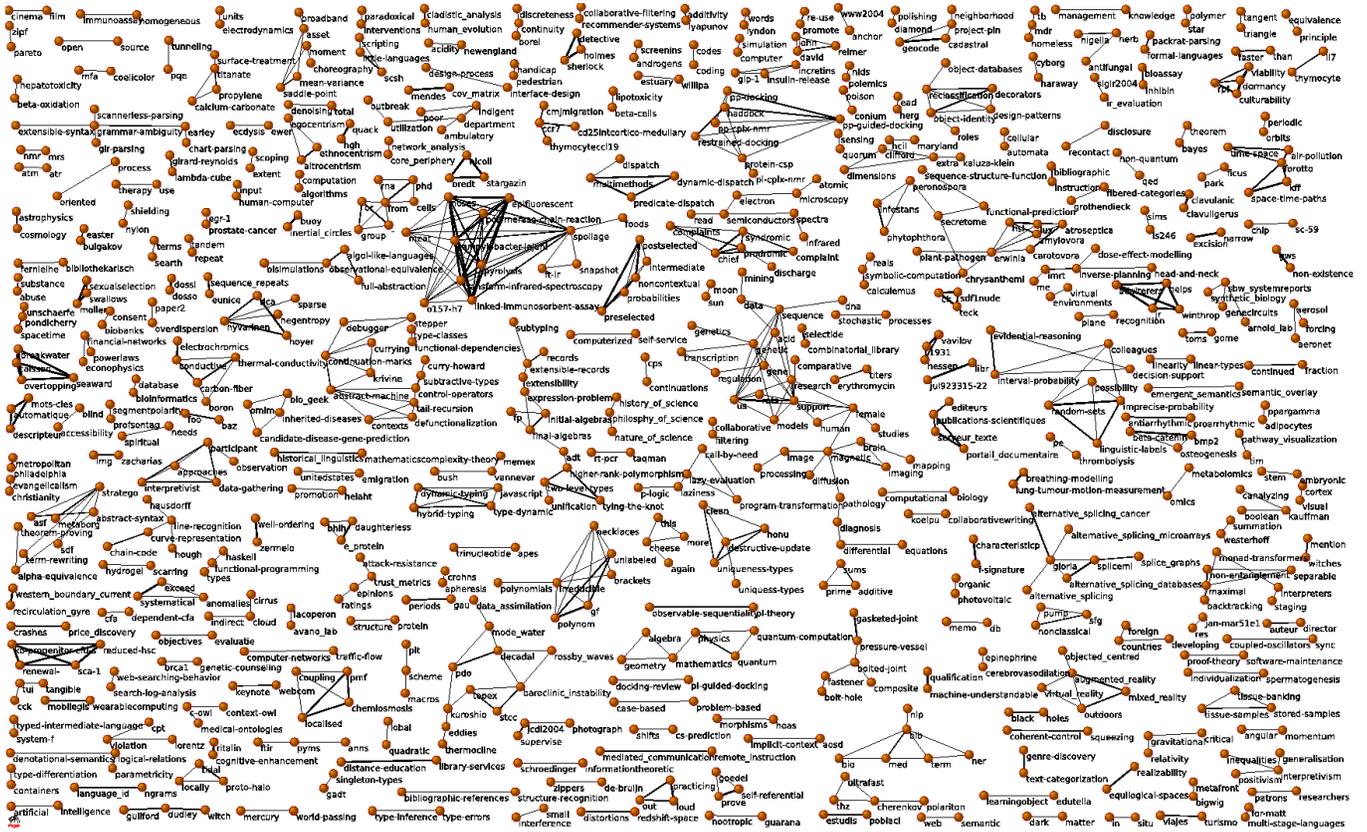}}}}
\caption{
The network of  tags in CiteUlike, constructed via the application of the similarity measure shown in Eqn.~\ref{eq:Similarity}---in this case considering the set of papers as neighbors.  Vertices with distance larger than $0.9$ have been discarded. 
The connection between tags ``differential'' and ``equation'' indirectly implies the clustering of the corresponding papers into the same group.
in the same ``category''}
\label{Fig:Similarity} 
\end{figure*}

Since most of the datasets (of significance) available from both Flickr and CiteUlike are fairly large
(on the order of a million hyperedges at the very least), we employ an approach based on  a
local quantity---vertex similarity. 
The vertex similarity is a measure of vertex ``distance'' defined as 
\beq\label{eq:Similarity}
\rho(v_1,v_2)=\frac{(N_1\cup N_2)-(N_1\cap N_2)}{(N_1\cup N_2)+(N_1\cap N_2)},
\eeq
where $N_1$ and $N_2$ are neighbors of the vertices $v_1$ and $v_2$ respectively.  The numerator is
the standard Euclidean---or Hamming distance in information theory---and the denominator is the
just the sum of the degrees.  Note that the measure can meaningfully be employed only to vertices of the same type (similar tags for example), and not necessarily to those of unlike types.  In addition, for tripartite graphs, one has to chose the type of neighbor.  So, if we were to look at the similarity between two tags, we can either consider its set of neighboring users or neighboring resources.
This approach is particularly useful for social networks like folksonomies, since it leads to self-categorizaton of content, via a bottom-up procedure (which seems natural for such decentralized systems).

In Fig.~\ref{Fig:Similarity} we show the result of the application of this to a subset of the data taken from CiteUlike. The figure shows groups of similar tags, where we used the papers as the neighbor set---in other words two tags are similar if they have employed many times together to different papers.
In this particular example tags are connected if their distance $\rho(v_1,v_2)$ is lower than a given threshold, in this case $\leq 0.9$. In principle one can tune these connected structures by modifying the value of this threshold parameter.  As the figure shows tags such as \emph{differential} and \emph{equation} indeed are similar to each other in the sense considered here.

\section{Conclusions}

In this paper we have examined the structural properties of two new types of social networks, so called folksonomies, consisting of users applying descriptive tags to resources. In order to preserve this three-way relationship, we have represented this structure as a tripartite hypergraph, with a user-resource-tag triple representing a hyperedge.

We have define a number of topological quantities of interests, such as a variety of degree distributions, correlations, clustering, distance distributions as well as a simple metric for discerning community structure. We then empirically measured these quantities on data taken from subsets of our example networks, Flickr and CiteUlike. We find that these networks share a number of qualitative features with previously studied social networks such as the presence of fat tails in the statistical distributions of links, the small world property in terms of the distance between vertices, as well as a high degree of clustering.

We propose that the topological measures as well as the methodology proposed here can be used as a standard tool for measuring the properties of networks of a similar nature.   

\begin{acknowledgments}
The authors thank CiteULike and the EU project TAGORA for providing the data analyzed in this paper. In addition we thank Mark Newman for his valuable insight and comments. 
\end{acknowledgments}

\end{document}